\documentclass[twocolumn,showpacs,preprintnumbers,
amsmath,amssymb]{revtex4}

\begin{document}
\title{The Two-Component Model and Metallization of Van der Waals Crystals}
\author{V. N. Bogomolov}
\affiliation{A. F. Ioffe Physical \& Technical Institute,\\
Russian Academy of Science,\\
194021 St. Petersburg, Russia}
\email{V.Bogomolov@mail.ioffe.rssi.ru}
\date{\today}

\begin{abstract}
  The paper discusses a model of Van der Waals crystals in which
band-gap structures do not form. An effect of strong and chaotic
electron-electron  repulsion, which was excluded from
consideration in the traditional approach, is taken into account.
A condensate exists as a result of a dynamic equilibrium among
atoms acted upon by constant Van der Waals forces and periodically
forming and disappearing covalent bonding. One part of atoms is,
on the average, in the ground, and the other, in excited state, to
form diatomic virtual molecules. Treated in terms of this
pseudoclassical model, the interatomic distances, binding
energies, volumes, and pressures at which metallization, for
instance, of inert gases and hydrogen, sets in is described by
simple relations involving only two spectroscopic parameters of
atoms (molecules). Applying pressure to a VdW crystals transfers
it from the insulator first to a Bose superconductor,  and  after
that,  to  a  Fermi metal.  An empirical relation $T_c \sim
N^{2/3}$ between the superconductivity transition temperature
$T_c$ and the particles concentration $N$  in chalcogens under
pressure is considered as an example of such situation.
\end{abstract}
\maketitle
   The problem of metallization of Van der Waals (VdW) condensates is one of the oldest in condensed
matter physics. It originated in XVIII century from the assumption
that hydrogen can undergo metallization in transformation to a
solid \cite{1}. One of the latest attempts at obtaining metallic
hydrogen experimentally and estimating the necessary pressures was
undertaken in \cite{2}.

   The properties of VdW condensates are dealt with in a large number of theoretical studies based on
traditional band-gap concepts. Some of the complicated problems
generated in this approach  still remain unsolved \cite{3}. One of
the reasons that could account for these difficulties lies in that
VdW condensates are possibly not band-gap materials at all. One
may, however, conceive of a model of the condensate in which
atomic interactions, rather than being constant, appear and
disappear periodically between pairs of atoms over the whole
lattice, i.e., they are localized, both in time and in space. The
validity of this model is argued for by the experimental support
obtained by the Herzfeld criterion, which defines the VdW
condensate volume at metallization in a straightforward way
through atomic properties as the atomic or molar refractivity (the
molar volume should become equal to the gas-phase molar
refractivity) \cite{4}. This supportive evidence was obtained from
xenon metallization ($V_m =$10.2{\mathsurround=0pt
cm$^3/$mol,}\quad $P_m = $160 GPa) \cite{5}.

 This VdW condensate model with a flickering, pulsating coupling between neighboring atoms, which
originates statistically throughout the lattice, was used to
interpret the results of the metallization of Xe as evidencing its
transition to a superconducting state with a very high $T_c>$300 K
\cite{6, 7}. In \cite{5}, Xe metallization ( from optical
characteristics) was treated as resulting from band closure and a
transition to a conventional metal. The possibility that the
"metallization" of the VdW condensates could be accompanied by the
onset of superconductivity at a superhigh-temperature did not
become a subject of serious consideration because of the
traditional belief that VdW condensates are typical band-gap
substances.

  Due to its inherent simplicity, the VdW condensate model proposed in \cite{6, 7} not only is in accord with
Herzfeld's approach, but yields, even in a very rough
approximation, as simple expressions for the condensate binding
energies $q$ and the critical volumes $V_m$ and pressures $P_m$ at
metallization (through the equation of state \cite{8} ). The only
things one needs here are two spectroscopic parameters of atoms or
molecules, namely, the ionization potential$ E_0$ and the
excitation energy $\Delta E:$
$$
q = C(E_0 -\Delta E)\exp \{-1.73/[(E_0 /\Delta E)-1]\};
$$
\[P_m=KE_0^4; \qquad \qquad V_m = F/{\Delta E}^3.\]

The parameters $C=1.44,\; K=0.0074,\; F=5880$ for the inert gases
were derived from data available for Xe. For helium He and
hydrogen H\mathsurround=0pt$_2$, which have only one electronic
shell, the parameter $C = 1.44/8.$ Using the known values of $E_0$
and $\Delta E$ (in eV) for these gases \vskip0.25cm \noindent He
(24.8 ;19.8); \quad  Ne (21.56; 16.6); \quad Ar (15.76; 11.6);

\noindent   Kr (14.02; 9.92); \quad Xe (12.13; 8.32); \quad
H$_2$(15.43; 11.2), \vskip0.25cm \noindent one obtains the
following ratios of the calculated to experimental values $q/q_e:$

\vskip0.25cm {\noindent (a)\quad He (0.00098/0.00098); \quad Ne
(0.021/0.018);

 \quad  Ar (0.048/0.063); \qquad  Kr (0.089/0.090);

 \quad Xe (0.126/0.126); \qquad H$_2$(0.0080/0.0093),}\vskip0.25cm

\noindent and the volumes at metallization {\mathsurround=0pt
(cm$^3$/mol)}, $V_m,$ relative to the values by Herzfeld,
$V_{mH}:$

\vskip0.25cm {\noindent (b) He (0.76/0.50); \quad Ne (1.28/1.00);
\quad Ar (3.77/4.16);

\nopagebreak \quad Kr (6.02/6.22); \quad Xe (10.2/10.2); \quad {
\mathsurround=0pt H$_2$}(4.2/2.0), \vskip0.25cm

\noindent and the metallization pressures $P_m$ (GPa):

\vskip0.25cm {\noindent (c)\quad He (2800); \quad Ne (1600); \quad
Ar (457);

\qquad Kr (286); \quad Xe (160); \quad H$_2$(420).} \vskip0.25cm

 The metallization pressure of hydrogen is close to 450 \nolinebreak GPa, the estimate obtained in \cite{2}.
\mathsurround=4pt In the model under discussion, the VdW
condensate lattice exists due to a dynamic equilibrium between two
virtual "phases". One of them, the "ground-state phase", comprises
atoms in the ground state, which are "slowly"  brought together by
the constant forces of Van der Waals attraction. The band-gap
properties are not efficient (in accordance with the Mott
criterion) on account of the large interatomic distances, because
the atomic electron shells are filled. The principal quantum
numbers of the first free and last filled electronic levels in an
atom differ by one, and the diameters of the orbitals differ by
2--3 \AA. The other "phase" is actually a gas of diatomic virtual
covalent-bonded molecules, which form and break up in a random
manner within the ground-state phase of virtually excited atoms
(excimers). Because of disorder and short lifetimes of the virtual
states, band-gap structures do not form in this phase either. The
lattice is maintained in equilibrium due to particles transferring
continually from one phase to another. The fluctuations in
electronic system   $ \sim  (E_{0}$ - $\Delta E$)
\mathsurround=1pt eV  are followed by the lattice vibrations (mean
energy about ${(E_0 -\Delta E)m_e/M_a \sim 0.7}$ meV for He). A
new type of \mathsurround=4pt localized excitations ($\sim 1$ meV)
were observed in solid helium \cite{9}.

This pseudoclassical pattern reminds one of a garland of balls
suspended in air, which exists due to periodic pulses received by
the balls from the juggler's hands. Ideas that could be treated as
supporting such a VdW condensate scheme appeared after
Ref.\cite{4} as well. These are, for instance, the chemical bond
resonance theory of Pauling and detection of "virtual
Xe{\mathsurround=0pt $_{2}$ molecules"\cite{10} in solid Xe. The
enigmatical high mobility of hydrogen atoms in palladium may be a
result of resemblance the electronic structures of
Pd$(4d^{10}5s^{0})$} and atoms of inert gases and existence of the
ground state atom sublattice in ordinary space.

{The "juggler hand" mechanism producing virtual molecules is also
very simple. The wave functions of atoms in the first phase are
not coherent (there are no band gap structures). Therefore, the
strong and random electron-electron interaction   of average
energy $ \sim e^2/D$ between atoms is  similar to the thermal
vibration field in conventional condensates, where the excitation
probability of a single impurity $p\sim\exp(-\Delta E/kT).$
Equating this energy to the average thermal vibration energy
$3kT/2 = e^2/D$ yields for the probability of atom excitation in
the first phase
\tolerance=400}
 \[ X\sim \exp[- ( 1.5\Delta E D/
e^2 )] ,\quad (\Delta E D) = f (P).\] This is the probability for
an atom of the first phase to transfer to the second phase, or the
relative particle concentration in the second phase. The
interatomic distance $D$ can be expressed through the
hydrogen-like diameter of an excited atom, with due account of its
increase (1.15) in the condensate:
$$
                                  D = 1.15\, e^2 / (E_0 -\Delta E ).
$$
The diameter of atom in ground state is $d_0 = e^2/E_0$ and the
Mott relation $D >2d_0\quad  (0.5 > 1-(\Delta E/E_0 ))$ is valid
for inert gases and {\mathsurround=0pt H$_2.$}

   Assuming such excitations to form in pairs to produce virtual molecules, we obtain for the number of
covalent molecules $\sim X << 1,$ and for the number of atoms in
the ground state and with a weak Van der Waals attraction, $\sim(1
- X).$ Therefore the average binding energy of a VdW condensate $
                                  q\sim Xe^2 /D.
$

   It is these simple expressions that underlie the above relations (at metallization {$X_m$ and $\Delta E_m D_m$} are
const.), which describe in a consistent way the VdW condensate
properties. One can derive in a similar manner a relation for the
compressibility and the equation of state \cite{8}, which is in
agreement with the results reported in \cite{5}:

\medskip $-[ d V/ VdP ]\sim V / q;$

\medskip $P\sim E_0^4\int y^{-7 / 3} \{\exp[-1.5( y^{1/ 3}- 1)]\}dy;$

\medskip $y = V/v;\quad V^{1/3}\sim D; \quad v^{1/ 3} \sim e^2/ E_0.$

\medskip  Within this VdW condensate model, the only cooperative effect, as in the band model, is the establishment
of an average interatomic distance through the inertia of atoms,
that vibrate as a result of periodically forming and breaking-up
covalent molecules at fixed separations and of the Van der Waals
attraction. And it is only at "metallization" ($X_m\sim1$) that
cooperative effects of the electronic type become possible. It was
suggested that the VdW "metallization" is actually
superconductivity with a very high $T_c$ \cite{6, 7}. An electron
pair in a virtual molecule, for instance, in { \mathsurround=0pt
Xe$_2,$} may be considered as a tight-bound boson $(\sim0.5 $ eV),
a Cooper pair two lattice constants in size. Its bonding is
realized through a "two-atom lattice". At the VdW condensate
metallization, the concentration of such "molecular" small radius
bosons in a gas reaches the Bose condensation point. There is no
problem of boson formation or decay at real temperatures, as is
the case with metals . We are aware of only one attempt at
detecting superconductivity of Xe from conductivity
characteristics \cite{11}. In this experiment, however, one could
not raise the pressure above 155 GPa. The change in the color of
{\mathsurround=0pt H$_2$ under compression \cite{2}, as well as in
the compression of Xe at $P<P_{m}$ \cite{5}, could be due not to
band closure, but rather to the absorption of light by particles
of the forming metallic or Bose phase \cite{6,7,12}.} The
diamagnetic "regions" (undetectable in conductivity \cite{11})
were observed in HTSC materials at $T > T_c$ \cite{13}. It is well
known that the color of systems with conducting particles depends
on both the particle size and their concentration (recall, for
instance, the coloring of metal sols).

    It may be conceived that an electron pair in a divalent atom ("atomic" boson) embedded in a
metal, rather than breaking down completely because of the
delocalization of its electrons, as is the case with electrons in
univalent atoms, remains a "pseudomolecule", albeit very weakly
bound {($ E\ll E_{0})$}. At $T = 0,$ these "atomic" bosons, which
are not completely "dissolved" in the metal (the Cooper pairs),
form a Bose condensate to give rise to superconductivity, which
usually does not set in in the case of univalent atoms. The BCS
theory provides an explanation for what prevents the electrons of
such "relic" bosons in metals from complete delocalization. For
tight-bound "molecular", noninteracting bosons in a VdW condensate
or for "atomic" ("relic") bosons (or "relic" atomic wave
functions) in a dense gas of divalent atoms (Mg, for example),
such danger apparently does not exist (before metallization).

Divalent atoms or diatomic molecules of univalent atoms dissolved
in dielectric media, are diamagnetic. For instance, it is a
{\mathsurround=0pt Na$_2$ molecules in NH$_3$ ( but not an
"individual electrons self-trapped in physical cavities in solvent
medium") \cite{14}. A volume per atom in Na$_2$ molecule is larger
of that in metallic Na.}

    Despite the simplicity of the structure of the VdW condensates, their most intriguing property, namely, the
possibility of providing superhigh-temperature superconductivity
(with the atoms being nonmagnetic) remains a mystery. There is, an
experimental problem of increasing the boson concentration either
through an increase in pressure or by physicochemical means
\cite{14}. Metallic {\mathsurround=0pt Pd$(4d^{10;9}5s^{0;1})$}
may represent roughly a final state of the VdW condensate under
pressure. It is a paramagnetic $(+540\times
10^{-6}${\mathsurround=0pt cm$^{3}$/mol).} Nevertheless, a
compounds {\mathsurround=0pt PdH$_{0.6},$  PdH and (Pd$_{0.7}$
Ag$_{0.3}$)H$_{0.8}$}  are diamagnetic and superconductors (9 K
and 16 K) respectively \cite{15}.
\{tolerance=400
 Applying
pressure to a VdW condensate (a gas of noninteracting bosons)
transfers it from the insulator first to a Bose superconductor,
and after that, to a Fermi metal \cite{14, 8, 16.} An example of
such situation is, perhaps, behaviour of $T_c$ in chalcogens under
pressure \cite{17}. The phase transition {\mathsurround=0pt
(S$_{8}$ $\to$ S$_{6}$)} in sulphur (close to insulator) is
followed by increase \makebox {in concentration $N$
of molecules}\\
\smallskip\makebox[32em][l]{ ${(N_{6} \sim N_{8}\times(8/6)^{2})}$
and by increase in $T_c$}
\\\smallskip $T_{c6}/T_{c8} = 17.3/12.5 =1.38 \sim(N_6/N_8)^{2/3}
=1.46$).\\
Under pressure the molecular structures of  Se and Te (close to a
Fermi metal) disappears and the phase transition
{\mathsurround=0pt { (Se; Te)$_{6}$ $\to$ (Se; Te)$_{1}$ } is
followed by increase in concentration of
particles $(N_{1}\sim 6N_{6})$ and by increase in $T_c$:\\
\smallskip \makebox[15em][c]{ Se :  $T_{c1}/T_{c6} = 10.3/3.4 =
3.0;$}\\
\smallskip \makebox[15em][c]{Te :  $T_{c1}/T_{c6} = 7.5/2.5 =
3.0.$}
It is close to relation\\
\smallskip \makebox[18em][c]{$T_{c1}/T_{c6}\sim(N_1/N_6)^{2/3}\sim6^{2/3} =
3.3$} and do not contradict to the Bose condensation mechanism of
superconductivity. Relatively low $T_c$ may be a result of  the
virtual molecules paramagnetism.

    Besides the hydrogen and inert gases, there is a large number of VdW condensates with a relatively low
$P_m$  and, perhaps, with very high $T_c$ superconductivity
\cite{6, 7}.

   The VdW condensate is an  example of an intermediate state of matter between a gas and a solid, between
micro-and macroobjects with combined electron systems. The
interactions in system are below a threshold value and  band-gap
structures (coherent state) do not form. It is the origin of
difference between the chemical substances and the sorption
compounds \cite{18}. Such materials can be constructed of atoms
with either filled or unfilled electron shells. Properties (a),
(b), and (c) can be readily explained in terms of both the model
under discussion and the band approximations. Only the concepts of
VdW condensate "metallization" are quite different.

\end{document}